\documentclass[prd,twocolumn,floats,floatfix,showpacs,nofootinbib]{revtex4}
\usepackage{graphicx}
\usepackage{dcolumn}
\usepackage{bm}
\usepackage{graphics}
\usepackage{slashed}
\usepackage{amssymb}
\usepackage{natbib}
\usepackage{amsmath}
\usepackage{amsthm}
\usepackage{url}
\usepackage{color}
\newcommand{\bea}{\begin{eqnarray}}
\newcommand{\Z}{\alpha +\beta}
\newcommand{\T}{{x^0}}
\newcommand{\ena}{\end{eqnarray}}
\newcommand{\bean}{\begin{eqnarray*}}
\newcommand{\enan}{\end{eqnarray*}}

\newcommand{\fracc}[2]{\frac{\textstyle{#1}}{\textstyle{#2}}}

\newcommand{\0}{t}
\newcommand{\B}{2 \sqrt A}
\begin{document}
\newcommand{\bb}{b}
\newcommand{\tr}{\textcolor{black}}
\hyphenpenalty = 10000

\title{Cosmology in GSG}

\author{E. Bittencourt$^1$} 
\author{U. Moschella$^2$} 
\author{M. Novello$^3$}
\author{J.D. Toniato$^3$} 

\affiliation{$^1$CAPES Foundation, Ministry of Education of Brazil, Bras\'ilia, Brazil and\\
Universit\`a di Roma La Sapienza - Dipartimento di Fisica\\
P.le Aldo Moro 5 - 00185 Rome - Italy}

\affiliation{$^2$Universit\`a degli Studi dell'Insubria - Dipartimento DiSAT\\
Via Valleggio 11 - 22100 Como - Italy and\\
INFN, Sez di Milano, Via Celoria 16, 20146, Milano - Italy}

\affiliation{$^3$Instituto de Cosmologia Relatividade Astrofisica ICRA - CBPF\\
Rua Dr. Xavier Sigaud 150 - 22290-180 Rio de Janeiro - Brazil\\}

\date{\today}

\begin{abstract}
We describe what cosmology looks like in the context of the geometric theory of gravity (GSG) based on a single scalar field. There are two distinct classes of cosmological solutions. An interesting feature is the possibility of having a bounce without invoking exotic equations of state for the cosmic fluid. We also discuss cosmological perturbation and present the basis of structure formation by gravitational instability in the framework of the geometric scalar gravity.
\end{abstract}

\maketitle
\begin{sloppypar}

\section{Introduction}\label{intro}

Although General Relativity (GR) is incontestably the paradigm
for the study and the description of gravitational phenomena, many alternative proposals
 have been put forward in recent years. These attempts try to modify GR in several respects, in particular
 to get around the need of introducing dark matter and dark energy which represent things
unknown at the laboratory scale.

Most of the proposed
models are effective in the realm of cosmology, where the intensity of
the gravitational field is \tr{strong} enough to excite new phenomena, possibly not contemplated by GR.
However, \tr{because of} the highly
paradigmatic status of GR, only a narrow
set of possible modifications  has been discussed to date.
\tr{A typical example consists in replacing  the scalar curvature $ R $ in the Einstein-Hilbert action}
with a somehow arbitrary
function $ f(R).$

\tr{This is not the road that one should follow according to a certain time-honoured thinking of the epistemological critique.
Following Mach, the natural way to undertake a deep modification of a paradigmatic theory is to
return to the main ideas of its anlagen and reexamine its foundations}
and  its evolutionary way to become paradigmatic in a historical context.
Alternative theories can emerge in this way.
Of course this is not a guarantee of success for a new proposal
that intends to substitute or modify substantially the established theory.
But at least it is a road that a scientist \tr{may want to} follow
to find a realistic alternative to the ancient scheme.

After the advent of special relativity it became evident that a profound
modification of Newton\rq s gravity was
unavoidable. The simplest one was to trade
the unique \tr{spatial} Newtonian potential $ \Phi_{N}$  for a
scalar field $\Phi$ defined on the full Minkowski spacetime.
\tr{However, the proposals by Nordstr\"om \cite{nord}, Einstein and Grossmann \cite{entwurf}   and a few other similar attempts were not successful
and rapidly discarded.} The difficulties of  those scalar theories and the reasons for their dismissals
are described e.g. in \cite{deruelle},
\cite{giulini} (and references
therein).
Drawbacks range from theoretical to observational; they \tr{are consequences} of
the hypothesis already stated in the Einstein-Grossmann
proposal \cite{entwurf} that scalar gravity \tr{should emerge from}
three \tr{basic} assumptions:
\begin{itemize}
\item{The theory is described in a conformally flat geometry and the background
Minkowski metric is observable;}
\item{The source of the gravitational field is the trace of the
energy-momentum tensor;}
\item{The scalar field is the (special) relativistic generalization of the Newtonian potential.}
\end{itemize}

In a recent paper \cite{JCAP} \tr{we have proposed an alternative way} for describing the
gravitational interaction in terms of a scalar field $\Phi$,  a
geometric scalar gravity (GSG), where the above three assumptions do not hold:
{in particular, in GSG the source of the gravitational field is not the trace
of the energy-momentum tensor and the theory is not a special
relativistic scalar gravity}. The adjective geometric
pinpoints indeed that GSG is a metric theory of gravity, as GR is.
We follow
the main idea of general relativity and assume as an \textit{a
priori} that gravity is described by a \tr{Lorentzian} curved geometry. In
general relativity the ten components of the metric tensor are the
basic variables of the theory (up to diffeomorphism invariance).
Here the metric tensor is determined by the gradient of one
fundamental independent physical quantity represented by the scalar
field $ \Phi.$ This means that although we make use of a scalar field to represent
gravitational interaction, we do not follow the previous examples
of scalar gravity, as for instance the Einstein-Grossmann ``Entwurf
Theory" \cite{entwurf}. We argue that \tr{most of the}
difficulties of the previous
scalar models
are rooted in their missing the main
idea that makes the success of general relativity
 that is {\em gravity is a metrical phenomenon}.

In other words we may recognise in a metric theory of gravity two separate
independent assertions:
\begin{itemize}
\item{gravity is a geometrical phenomenon represented by a \tr{Lorentzian} metric $ g_{\mu\nu}$;}
\item{the quantity $ g_{\mu\nu}$ may be associated either to a symmetric
tensor $ h_{\mu\nu}$ or to a scalar $\Phi$, \tr{or some other field possibly depending on tensor indices.}}
\end{itemize}
Einstein \tr{chose} to introduce a full second order tensor to represent
the gravitational metric. However this \tr{was not}
mandatory. The above two assertions are formally independent, and the first one comes first.
\tr{The choice between the different
possibilities mentioned in the second statement  that produce acceptable models  (if any)
in agreement with observations above is a
matter of taste and mathematical ability}.

Several \tr{objections} have been raised against scalar gravity (see
for instance \cite{feynman}). A \tr{careful} inspection \tr{shows that those criticisms}
apply to models that do not satisfy the first (Einstein's) hypothesis and
do not relate scalar gravity to a modification of the
space-time geometry; once more,this is the \tr{main} drawback of the old
proposals of scalar gravity.

All these criticisms do not apply to GSG.
\tr{Furthermore, our simple model \cite{JCAP} gives right answers - the same as in GR -
at least for some important gravitational phenomena such as the solar system tests and at the cosmological level. GSG also provides a geometry for slowly rotating bodies (as those observed in gravity probe-B experiment) similar to the one found in GR}.

\tr{How it is that} such a simple theory - that uses only one function to
describe gravity rather than ten -- can account
for all these phenomena along the same lines as GR?
Should this fact points in favour of a \tr{feature} hidden in GR that
GSG is \tr{capturing} or is it possible
that such GSG \tr{(or a more refined version of it)} can effectively \tr{produce a} new and
independent view of gravitational processes?

Irrespective of the answer to these questions,
there are certain properties of GSG that \tr{may}
solve certain problems that GR is able to deal with only making appeal to unusual behaviour of matter.
\tr{In particular, in this paper, we show how GSG gets around the question of the
cosmological singularity that is understood as almost inevitable in
standard relativistic cosmology, but that is avoided in GSG as we
are going to show.}

Other features that deserve to be mentioned here are the possibility offered by GSG to define
an energy-momentum tensor for gravity
as opposed to the pseudo-tensors that have been proposed to represent the energy
distribution of gravity in GR.
Also, the dynamical bridge
relating the propagation of the scalar field in an unobserved auxiliary
Minkowski space-time and the corresponding dynamical propagation in
the gravitational metric, might suggest that perhaps it
should be simpler to quantize gravity using ideas from GSG.

In the end of this introduction we would like to recall that GSG is newly born and its properties and solutions are largely yet unknown. \tr{For instance, we still did not analyse completely the problem of radiation emission from compact systems, that is the main pillar sustaining the glory of GR}. Other aspects of gravity should be analyzed carefully in the future as well. \tr{We hope that this work can at least stimulate a regain of interest and thinking about scalar theories of gravity.}
\section{Geometric Scalar Gravity: a short summary}
The modifications of Newton\rq s gravity based on a scalar field,
proposed in the first decade of the 20th century, just after the birth of special relativity and before the advent of general relativity, all failed.

Some of the reasons behind those failures have been reported in \cite{JCAP}, where  we proposed an alternative model, also based on a scalar field,
that does not share the drawbacks of the old attempts. Let us recall its construction in a direct comparison with GR.
\subsection{Basic features of GR}
\begin{itemize}
\item A general metric  may always be decomposed
as the sum of a reference Minkowski metric plus a (not necessarily small) perturbation as follows:
\begin{equation}
g^{\mu\nu} \equiv \eta^{\mu\nu}+ h^{\mu\nu}. \label{9julho12}
\end{equation}
This binomial form is exact.
The covariant inverse
$g_{\mu\nu}$ is however, generally speaking, not a binomial but  an
infinite series:
$$ g_{\mu\nu} = \eta_{\mu\nu} -  h_{\mu\nu} + h_{\mu\alpha} \,
h^{\alpha}{}_{\nu} + ... $$
\item{The
second order tensor field $h_{\mu\nu}$ describing the gravitational interaction satisfies a nonlinear field equation. }
\item{The theory satisfies the principle of general covariance.
In other words, $h_{\mu\nu}$ is not a field restricted to the realm of
special relativity.}
\item{Newton's gravity is reproduced in a suitable approximation. $h_{\mu\nu}$ is related in a nontrivial way to the Newtonian potential $\Phi_N$.}
\item{The background Minkowski metric
is not observable.  Matter and energy interact gravitationally only
with $g_{\mu\nu}$. Test particles move along timelike
geodesics and
electromagnetic waves propagate along null geodesics relative to $ g_{\mu\nu}.$}
\end{itemize}

\subsection{Basic features of GSG}
\begin{itemize}
\item In GSG, Eq.\ (\ref{9julho12}) is specialised to a particular form built
in terms of a scalar field  $\Phi$:
\begin{equation}
 q^{\mu\nu} = \alpha \, \eta^{\mu\nu} + \frac{\beta}{w} \, \partial^{\mu}\Phi \,\partial^{\nu} \Phi
\end{equation}
where $ w=\eta^{\mu\nu} \, \partial_{\mu} \, \Phi \, \partial_{\nu} \Phi$; the  covariant inverse  is
now also a binomial:
\begin{equation}
q_{\mu\nu} = \frac{1}{\alpha} \, \eta_{\mu\nu} -
\frac{\beta}{\alpha \, (\alpha + \beta) \, w} \, \partial_{\mu} \Phi
\, \partial_{\nu} \Phi.
\label{9junho11}
\end{equation}
\item{The
scalar field $ \Phi$ describing the gravitational interaction satisfies a nonlinear field equation. The functionals $\alpha$ and $\beta$ depend on $\Phi$ and are fixed according with observations.
 Use of the solar system tests and the Newtonian limit   \cite{JCAP} leads to the choice
\begin{eqnarray}
&& \alpha = \exp({- 2 \,\Phi}), \label{pota}\\ && Z= \alpha+\beta=\alpha^3 V=\frac{(\alpha-3)}{4}^2. \label{pot}
\end{eqnarray}}
\item{The theory satisfies the principle of general covariance.
In other words, $\Phi$ is not a field restricted to the realm of
special relativity.}
\item{Newton's gravity is reproduced in a suitable approximation.  $\Phi$ is related in a nontrivial way to the Newtonian potential $\Phi_N$.}
\item{The background Minkowski metric
is not observable. All kind of matter and energy interact with $\Phi$
only through the pseudo-Riemannian metric
$q^{\mu\nu}$. Test particles follow timelike geodesics
and electromagnetic waves propagate along null geodesics relative to the metric $ q^{\mu\nu}.$}
\end{itemize}

\subsection{Field equations}
In GSG the space-time metric is not a fundamental independent quantity
but a function of the scalar field $\Phi$. The latter is assumed to satisfy the following equation  (see \cite{JCAP} for details):
\begin{equation}\label{dynamic}
    \sqrt{V}~\square\Phi=\kappa \,\chi,
\end{equation}
where $\kappa=8\pi G/c^4$. In this equation $V$ is the potential given in Eq.\  (\ref{pot});
the source of the field at the rhs is constructed out of the following expressions:
\begin{eqnarray}\label{rhs}
E=\frac{1}{\Omega}~T^{\mu\nu}~\partial_\mu\Phi~\partial_\nu\Phi\\
C^\lambda=\frac{\beta}{\alpha\Omega}\left(T^{\lambda\mu}-Eq^{\lambda\mu}\right)\partial_\mu\Phi\\
\chi=-\frac{1}{2}\left[T +\frac{\alpha+3}{\alpha-3}~E + C^\lambda_{~;\lambda}\right]\label{rhs3}
\end{eqnarray}
where $T=q^{\mu\nu}T_{\mu\nu}$ and
$\Omega=q^{\mu\nu} \partial_\mu \Phi \, \partial_\nu\Phi $.
\section{Cosmology}
\subsection{Vacuum cosmological solutions}
Let
$ds_M^2=\eta_{\mu\nu}dx^\mu dx^\nu$
be the auxiliary Minkowski metric written in the standard way and suppose that the scalar field depend only on the time coordinate $\T$ i.e.
$\Phi=\Phi(\T).$
The nonzero coefficients of $q^{\mu\nu}$ are simply
\begin{equation}
q^{00} = \Z, \ \  q^{11} = q^{22} = q^{33} = - \,\alpha,
\end{equation}
and the line element takes the form (see Eq.\ \ref{pot})
\begin{eqnarray}
ds^{2} &=& \frac{4 d\T^{2}}{(3-\alpha)^2}\,  - \frac{1}{\alpha} \, (dx^{2} + dy^{2} + dz^{2}).
\label{lineel}
\end{eqnarray}
In GSG homogeneity of the field forces isotropy of the metric.
Let us consider in the above circumstances the  vacuum field equation
\begin{equation}\label{dyn}
\square\Phi=0
\end{equation}
is promptly reduced to a quadrature:
\begin{eqnarray}&& \frac{\alpha-3}{2 \alpha^\frac 52 }\frac{\partial\alpha}{\partial \T} = \B, \label{pipo}\\[2ex]
 && {\frac{1-\alpha}{ \alpha^\frac 32 }}= \B\T \end{eqnarray}
where $A$ is an integration constant.
Eq.\ (\ref{pipo}) implies that the derivative of $\alpha(\T)$ diverges at $\alpha = 3$ (apart from the trivial case $A=0$).
Let us restrict for the moment our attention
to values of the potential where the Newtonian limit of the theory
takes place i.e. $0<\alpha<3$; the relation between $\T$ and $\alpha$ is monotonous and can be inverted by solving a simple cubic equation.
Let us denote the result
$\alpha=\alpha_0(\T)$ with the geometry given by Eq.\ (\ref{lineel}).
Here is a first unexpected result. In GR, the cosmological hypothesis forbids the existence of curved cosmological vacuum solutions (the Milne universe is nothing but the flat Minkowski space in disguise). However, this becomes possible going beyond the isotropic case. The well known Kasner solutions constitute a class of geometries representing empty spatially homogeneous but anisotropic universes. In GSG there exist spatially homogeneous and isotropic solutions representing empty conformally flat universes.

Let us examine Eq.\ (\ref{lineel}) 
a little closer. There are two apparent singularities
at $\alpha_0 = 0$ and  $\alpha_0 = 3$. Are they real singularities? Computing for instance the scalar curvature we get \begin{equation}
R 
=  9 A \, \alpha_0^3
\end{equation}
so that $R$ is perfectly regular at both ends of the interval $(0,3)$.
To understand better the situation it is advantageous
to introduce a new time variable by the following definitions:
\begin{equation}
t  =  \frac{1 }{ 3 \sqrt A} (\alpha_0(\T))^{-\frac 32 }, \ \ \ \alpha_0(t) = \alpha_0(\T(t) )= \left({3 \sqrt A t}\right)^{-\frac 23} \label{change}
\end{equation}
(where we supposed that $A>0$). Since
\begin{eqnarray}
\frac{\partial t }{\partial \T} =  \frac{1}{ \B} (\alpha_0(\T))^{-\frac 52 }  \frac{\partial \alpha_0 }{\partial \T} =
\frac{2}{3-\alpha_0(\T) } \label{change1} \end{eqnarray}
the new variable may be interpreted as the cosmic time and the metric takes the cosmological form
\begin{eqnarray}
ds^{2} &=&dt^2\,  -  \left(\sqrt {9 A}  t\right)^{\frac 23}\, (dx^{2} + dy^{2} + dz^{2}). \label{vacuumsol}
\end{eqnarray}
While the cosmic time $t$ initially belongs to the interval $({2}/(9 \sqrt{3 A} )<t< \infty)$, the solution (\ref{vacuumsol}) makes sense in the whole interval $(0,\infty)$.  The Ricci scalar is now written as follows:
\begin{equation}
R = \frac{2}{3t^2}.
\end{equation}
At $t=0$ there is an initial singularity {\em aka} a big bang. After the big bang an empty curved space-time comes into being.

\subsection{Cosmological equation}
By following  the previous example let us adopt the cosmic time $t$ in the line element (\ref{lineel}) by a similar change of coordinates:
\begin{eqnarray}
&& dt=2d\T/(3-\alpha ), \ \ \ \ \alpha=1/a(t)^2, \label{def_t}\\
\, \cr
ds^{2} &=&  \frac{4}{(3-\alpha)^2}\, d\T^{2} - \frac{1}{\alpha} \, (dx^{2} + dy^{2} + dz^{2}), \\[1ex]
&=& dt^{2} - a(t)^{2} \, (dx^{2} + dy^{2} + dz^{2}).
\label{him}
\end{eqnarray}
The relation (\ref{pota}) now reads $a = \exp \Phi$ and we see that the derivative of the field w.r.t. the cosmic time is interpreted as the Hubble parameter:
\begin{equation}
H= \frac {\dot a} a = \dot \Phi .\label{Hubble}
\end{equation}
The lhs of the field equation is rewritten as follows:
\begin{equation}\label{lhs}
    \sqrt{V} \,\square\Phi= a \, |3a^2-1|\, \left(\frac{\ddot a}{2a}+\frac{\dot a^2}{a^2} \right).\
\end{equation}
As for the rhs we used (\ref{def_t}) into (\ref{rhs3}) to obtain
\begin{equation}
\label{rhs1}
\chi=-\frac{1}{2}\left[T +\left(\frac{1+3a^2}{1-3a^2}\right)E + C^\lambda_{~;\lambda}\right].
\end{equation}
Because the gravitational field depends only on $t$, it is natural to expect that all the relevant quantities have only temporal dependence too.
For the sake of simplicity we proceed here with that cosmological hypothesis i.e. we assume that the components of the energy-momentum tensor of a perfect fluid decomposed in terms of the cosmic observers $v^{\mu}=\delta^{\mu}_0$ depends only on time. Then, one has immediately that
\begin{eqnarray}
&&E= T^{00} \equiv \rho, \quad T=\rho-3p,\nonumber\\[1ex]
&&C^0=0, \quad \mbox{and} \quad C^i=\frac{\beta}{\alpha H} T^{i0}.
\label{c1}
\end{eqnarray}
The divergence of $C^\lambda$ vanishes and  equation (\ref{dynamic}) reads
\begin{equation}\label{jr}
  {a}|3a^2-1|\left(\frac{\ddot a}{a}+2\frac{\dot a^2}{a^2} \right) = \kappa\left[3p +\left(\frac{2\rho}{3a^2-1}\right)\right].
\end{equation}
There are therefore two regimes classified by the sign of the quantity $3a^2-1$.
As we will see below, the time evolution respects that sign and therefore GSG cosmologies based on a perfect fluid are divided into two classes. We call the solutions belonging the first class (i.e. solutions such that $3a^2>1$) Big Universes and the solutions belonging to the the second one Small Universes.  The adjective ``small" makes allusion to the fact the scale factor takes values in a compact interval, but also when this occurs  the spatial section is of course infinite (and flat). The existence of two classes of solution is  a consequence of the choice of the potential (\ref{pot}). Note that despite of the line element, the dynamics of the theory is not invariant under rescalings $a(t)\rightarrow\mu a(t)$, where $\mu$ is a constant. This will have important consequences afterwards.
\section{Big Universes. The Bounce.}
In this section we assume that $3a^2>1$. We will see that  this inequality is conserved by the time evolution and therefore the universe bounces at a minimal value of the scale factor. Before doing this let us comment on the general structure of the cosmological equation (\ref{jr}).
When the acceleration is small compared to the squared velocity the above equation reduces to an equation
similar to the Friedmann equation:  in the case of a dust-like fluid we get
\begin{equation}
 \left(\frac{\dot a^2}{a^2} \right) \simeq f(a) { \rho }.
\end{equation}
where $f(a)$ is positive. In other cases deviations from GR are stronger. In particular in the opposite regime, when the squared velocity is small w.r.t. the acceleration, the above equation mimics the Raychaudhuri equation but with a sign of the rhs which is opposite w.r.t. GR:
\begin{equation}
 \frac{\ddot a}{a} \simeq \left[2 f(a) \rho + 3p\right].
\end{equation}
\subsection{Barotropic fluids}
For a barotropic  perfect fluid  whose equation of state is $p=\lambda\, \rho$ 
conservation of the energy-momentum tensor implies as usual that
\begin{equation}
    \rho=\rho_0\, a^{-3(1+\lambda)}.
\end{equation}
For big universes the cosmological equation becomes
\begin{equation}\label{eqnlin}
  \frac{\ddot a}{a}+2\frac{\dot a^2}{a^2}= \kappa \, \rho_0 \, \frac{2 + 3\lambda \,(3a^2-1) }{a^{4+3\lambda} (3a^2-1)^2}.
\end{equation}
This equation can be linearized by posing $u = \dot a^2$:
\begin{equation}\label{eqlin}
\frac{d }{da}  (a^4u) - \frac {2\kappa \rho_0 (2+3\lambda (3 a^2-1)) a^{1-3\lambda}}{(3a^2-1)^2}=0
\end{equation}
and easily integrated to give
\begin{eqnarray}
u=\dot a^2=\frac{A}{a^4} -  \frac{2\kappa\rho_0}{a^{2+3\lambda}(3a^2-1)}. \label{yyy}
 \\ \label{eqnlinacc}{\ddot a}=-{\frac{2A}{a^5} +  \frac{\kappa\rho_0(2+3\lambda)}{a^{3+3\lambda}(3a^2-1)}}+ \frac{6\kappa \, \rho_0  }{a^{1+3\lambda} (3a^2-1)^2}.
\end{eqnarray}
$A$ is a strictly positive constant of integration.
As it should be expected, the above expression is singular at  $3a^2 =1$.
However this value is unattainable because the square of the velocity becomes zero at a minimal value $a_m$  strictly grater than $1/\sqrt{3}$. At that point the universe bounces.

\begin{figure}
\begin{center}
\includegraphics[width=8.5cm,height=9cm]{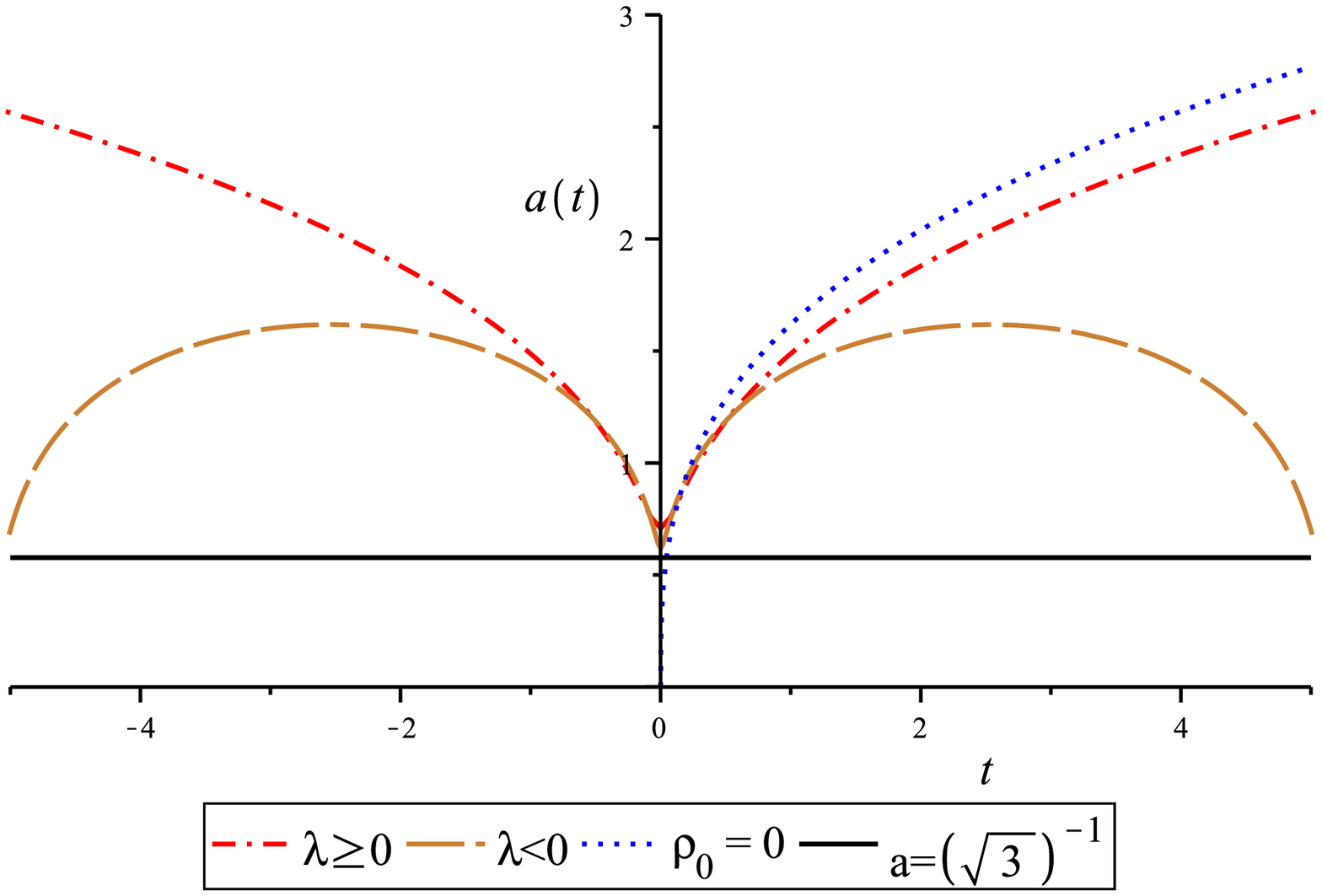}\quad
\includegraphics[scale=0.45]{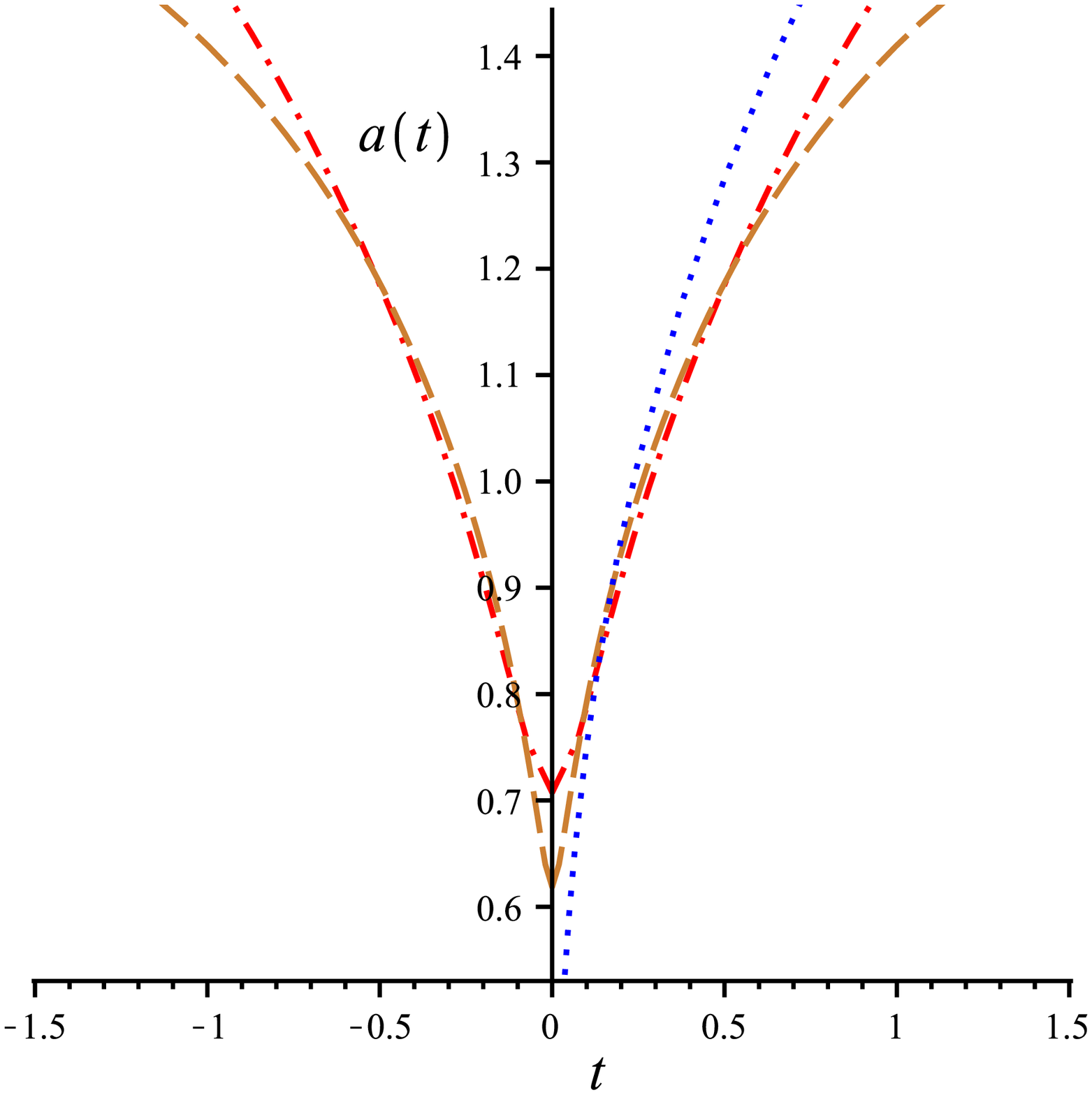}
\end{center}
\caption{(color online). On top the scale factors for $\lambda\geq0$ (dot-dashed red), $\lambda<0$ (dashed gold) and the vacuum solution $\rho_0=0$ (dotted blue), where we set $A=2$ and $\kappa\rho_0=1$ except for the vacuum case. The singularity $a=1/\sqrt{3}$ is represented by the horizontal solid line. On bottom, their behaviour in the vicinity of the bounce. Note that for the same values of the constants, the bounce for cyclic universes occurs at smaller values of the scale factor than the non-cyclic ones.}
\label{fig_bu}
\end{figure}

The possibility of having a bouncing for standard fluids is quite remarkable [see Fig.\ (\ref{fig_bu})].
In FLRW cosmology the bouncing is possible either by non-minimal
coupling with matter fields or by negative pressures.
Indeed, in order to have an extremum of the scale factor $a(t)$ the expansion factor $\Theta= 3  \dot{a}/a$ must vanish and its derivative be positive.
As it is well known, in GR the Raychaudhuri equation implies that this is possible when
$$ \dot{\Theta} + \frac{\Theta^{2}}{3} = - \, \frac{\rho + 3 p}{2}>0. $$
Here the situation is different: the universe always bounces with only one
notable exception the vacuum solution discussed before where there is an initial singularity.
Indeed, for empty universes (i.e. $\rho_0 \to 0$) Eq.\ (\ref{yyy}) reduces to
\begin{equation}
\dot a^2= \frac{A}{a^4} \label{empty}
\end{equation}
which is promptly integrated to find again that
\begin{equation}
a(t) = (\sqrt {9 A }\,  t )^\frac 13, \label{lt}
\end{equation}
apart from an irrelevant additive constant.

Letting $A\to 0$ in Eq.\ (\ref{empty}) we get as a special case the constant solution $a(t) = const$,
which represents a flat geometry. This solution has however to be intended as a limiting case.

\subsection{Dust}
Let us discuss the  case of dust in some detail. Eq.\ (\ref{yyy}) reduces to
\begin{equation}
u=\dot a^2= \frac{A}{a^4} -  \frac{2\kappa\rho_0}{3a^4-a^2} \label{dust0}
\end{equation}
Here the integration constant has to satisfy the following bound for the universe to exist:
\begin{equation}
A_0 = A-  \frac{2\kappa\rho_0}{3}>0.
\end{equation}
Eq.\ (\ref{dust0}) may rewritten in terms of the new constant $A_0$ as follows:
\begin{equation}
 \dot a^2=
 \frac{A_0}{a^4} - \frac{2  \kappa \rho_0}{3a^4}
 \frac{1}{(3a^2-1)} \label{dust1}
\end{equation}
At late times the universe filled with dust behaves as if it were empty,
i.e. $a(t) \sim ( \sqrt {9A_0}\, t)^{1/3}.$
On the other hand the cosmic evolution at early times is very different w.r.t. the empty case
(\ref{empty}).  As we already said the biggest difference is that there is no initial singularity but rather a bouncing at the minimal value of the scale factor
\begin{equation} a_{b}^{2} = \frac{1}{3} \, \left( 1 + \frac{2 \, \Delta}{3}\right) \label{ppp}
\end{equation}
where $ \Delta = \kappa \, \rho_{0} /A_0.$
At the bounce the acceleration is positive and inversely proportional
to the energy density $\rho_0$:
\begin{equation} \ddot{a}_{b}
= \frac{243 A_0^{7/2}}{2 \kappa  \rho_0  (3 A_0+2 \kappa  \rho_0 )^{3/2}}
= \frac{243 A_0}{2 \Delta (3 +2 \Delta)^{3/2}}
\end{equation}
Taking the limit $\rho_0\to 0$  (i.e. the vacuum limit) the minimal value of the scale factor $a_b$ tends to $1/\sqrt 3$ but the corresponding
 acceleration diverges;
there is however no contradiction with the regularity of the vacuum solution (\ref{vacuumsol}) because the dust solution
does not converge uniformly to the vacuum solution in this limit.

The accelerating phase does not last forever. The acceleration diminishes with the expansion and vanishes when the scale attains the value
$$ a_{d}^{2} = \frac{1}{3} \, \left( 1 + \frac{\Delta}{2} + \frac{\sqrt{\Delta^{2} + 4 \, \Delta /3}}{2} \right) $$
after which the universe starts decelerating.
On the other hand the scalar curvature
\begin{equation}
  R = - 6\left(\frac{\ddot{a}}{a} + \frac{\dot{a}^{2}}{a^{2}} \right)\label{10mar14}
\end{equation}
stays negative and changes signs only when the scale reaches the value
$$ a_{c}^{2} = \frac{1}{3} \,  \left( 1 + \frac{2 \, \Delta}{3} + \frac{\sqrt{2\Delta \, (2 \Delta + 3)}}{3} \right). $$

To put the results in perspective, in GR the scalar curvature is related to the trace of the matter energy-momentum tensor;
$R = -\kappa \, \rho$ in the case of dust. Thus, for pressureless fluid, no change of the sign of $R$ is possible.
In GSG the connection between $R$ and the energy content of the universe is more involved and the above example shows that such change is possible without  imposing  unusual conditions on matter content and/or on the geometry of the universe.

Fluids with positive $\lambda$ (including radiation) have a similar behaviour. They all have a bounce, followed by an early accelerated phase and a final decelerated phase. The expansion last forever.
At late times, they all share the same behaviour  (\ref{lt}) of the empty universe
irrespective of the value of $\lambda$.


\subsection{$\lambda<0$: static solutions and cyclic universes}
When the pressure is negative, i.e. for $\lambda<0$,  there exist
static cosmological solutions of Eq.\ (\ref{eqnlin}):
\begin{equation}
\label{size_stat}
a_0=\sqrt{\frac{1}{3}-\frac{2}{9\lambda}} > \frac{1}{\sqrt{3}}
\end{equation}
corresponding to the integration constant
\begin{equation}
A_\lambda =  3 |\lambda| \kappa\rho_0\,\left(\frac{1}{3}+\frac{2}{9|\lambda|}\right)^{1+\frac{3|\lambda|}{2}}
\end{equation}
Note that this is the minimal possible value of the constant $A$ relative to the equation of state $\lambda$ (below this value $\dot a ^2$ would be negative for all $a$).
For any choice $A>A_\lambda$ there are two values  where $\dot a = 0$. Since the acceleration is positive for $a = a_{min}$ and negative for $a= a_{max}$ the latter are inversion points and  the universe eternally oscillates  between a minimal and a maximal value of the cosmic scale factor
$a_{min} \leq a \leq a_{max}.$
Therefore a negative pressure produces a static universe which
at variance with GR is spatially flat and is stable.

\section{Small universes}
\begin{figure}
\begin{center}
\includegraphics[scale=0.45]{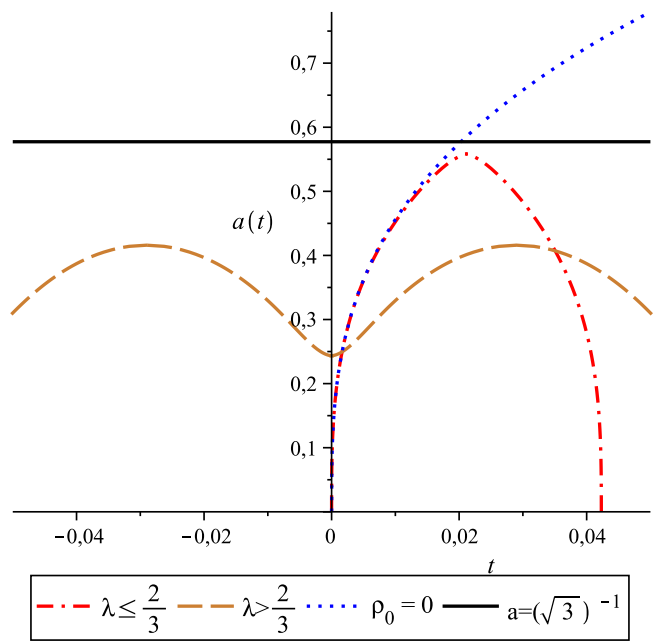}
\end{center}
\caption{(color online). The scale factors for $\lambda\leq2/3$ (dot-dashed red), $\lambda>2/3$ (dashed gold) and the vacuum solution (dotted blue). Here we set $A=10$. The solid black line is the singularity $a=1/\sqrt{3}$.}
\label{fig_su}
\end{figure}

We now shortly consider the small universes, i.e. cosmologies such that $a^2 < 1/3$ [cf. details in Fig.\ (\ref{fig_su})].
Taking into account the absolute value at the lhs of Eq.\ (\ref{jr}), for small universes the cosmological equation becomes
\begin{equation}\label{eqnlin2}
  \frac{\ddot a}{a}+2\frac{\dot a^2}{a^2}= - \kappa \, \rho_0 \, \frac{2 + 3\lambda \,(3a^2-1) }{a^{4+3\lambda} (3a^2-1)^2}.
\end{equation}
Proceeding as before
we solve for the scale factor:
\begin{equation}
\label{smallu}
\dot a^2= \frac{A}{a^4} -  2\kappa\rho_0\,\frac{a^{-2-3\lambda}}{1-3a^2}.
\end{equation}
Again, $A$ is a positive integration constant and, once more, the singularity at $a=1/\sqrt 3$ cannot be attained.
Therefore big universes and small universes are two disjoint classes of cosmological solutions of GSG.

Now static solutions exist for positive values of the parameter $\lambda$. More precisely for $\lambda>2/3$  there is an equilibrium solution at \begin{equation}
\label{size_stat2}
a_0=\sqrt{\frac{1}{3}-\frac{2}{9\lambda}} < \frac{1}{\sqrt{3}}
\end{equation}
corresponding to
\begin{equation}
A_\lambda =  3 \lambda  \kappa   \rho_0 \left(\frac{1}{3}-\frac{2}{9 \lambda }\right)^{1-\frac{3 \lambda }{2}} .
\end{equation}
For any choice $A>A_\lambda$ there are two values where $\dot a = 0$. The singularity at $a=0$ cannot be attained and the universe bounces at $a=a_m>0$. All in all, for $\lambda > 2/3$ and $A> A_\lambda$ the universe eternally oscillates between a minimal and a maximal value of the cosmic scale factor
$0\leq a_{min} \leq a \leq a_{max}<1/\sqrt 3$.

On the contrary, for\footnote{$\lambda = 2/3$ is a limiting case that we will refrain from discussing here} $\lambda <2/3$ the universe always runs into the singularity at $a=0$. In these cases there is an initial big bang followed from a decelerated phase. The universe reaches its maximum scale and then stops entering in contracting phase that unavoidably ends in the singularity at $a=0$.
It should be remarked that while such a behaviour is possible in FRW cosmology only for spatially closed universes, here we are always dealing with a conformally flat universe, i.e. the spatial section is flat.

{We can have some hints about the values of the parameters entering in Eq.\ (\ref{smallu})
by using the values of the cosmological parameters today (t). So, let us suppose that the baryonic matter contribution is the most relevant fluid at this time\footnote{This is an important starting point because we have to be sure whether the conventional fluids are enough to describe the observational data or not.}. Thus, we obtain
$$\frac{A}{a^6_t}=H_{t}^2+\frac{16\pi G\rho_t}{a_t(1-3a^2_t)}.$$
If the parameter $H_t=100h\,\rm{km\,s}^{-1}\rm{Mpc}^{-1}$ and $\rho_t\leq3\times10^{-10}\,\mbox{erg cm}^{-3}$ are not constrained by any model\footnote{Data collected from Ref.\ \cite{narlikar}.}, then it is a formula between $A$ and $a_t$. Another way to determine these arbitrary parameters is at the bounce. When $\dot a=0$, we get
$$A=\frac{16\pi G\rho_ba^5_b}{a_b(1-3a^2_b)},$$
where the cosmological parameters in this case are the size of the bounce $a_b$ and the energy density $\rho_b$ at that moment. Note that these formulas are only useful for a model-independent analysis of the cosmological parameters.}

\section{Cosmological perturbations} \label{cosper}
The aim of this section is to explore whether in GSG cosmology there is room for structure formation by the gravitational instability of initially small overdensities of an homogeneous fluid. We will focus on the behaviour of perturbations of dust cosmology in the Big
Universe sector and in the decelerated phase.

Linear scalar perturbations in GSG correspond to perturbations of the scalar field
$$\Phi\rightarrow\Phi+\delta\Phi,$$
where $\Phi$ is the background field and $\delta \Phi$ is a ``small" perturbation.
%

Let us  start by considering the scalar field as a function of $x^0$ where $\{x^\mu\}$ is a set of
Lorentzian coordinates; in the above coordinates the Minkowski metric is written in the standard form $\eta_{\mu\nu}=\mbox{diag}(1,-1,-1,-1)$.
At first order in $\delta \Phi$ the unperturbed metric gets the following corrections:
\begin{eqnarray}
\delta q^{00}&=& -\alpha(\alpha-3)\,\delta \Phi \label{p1}\\[1ex]
\delta q^{0i}&=& -\beta \delta^{ij}\frac{\delta\Phi_{,j} }{\Phi_{,0}} \label{p2}\\[1ex]
\delta q^{ij}&=& 2\alpha\,\delta\Phi\,\delta^{ij}\label{p3}
\end{eqnarray}
where the comma indicates partial derivative w.r.t. the corresponding coordinate, as usual. It is seen that the perturbation $\delta q^{0i}$ will not be small at all when ${\Phi_{,0}}\simeq 0$.
In that case the perturbative approximation may break down or else that problem may be just a gauge artefact.

Expressions (\ref{p1}-\ref{p3}) represent true perturbations of the gravitational field and they cannot be generated by any coordinate transformation (see Appendix \ref{A}). Thus, we refrain here to use the ideas and the full arsenal of gauge invariant perturbation theory \cite{mukhanov} (adapted to the present case).

Indices are lowered according to the relation  $$\delta q_{\alpha\beta} = - q_{\alpha\mu} q_{\beta\nu} \delta q^{\mu\nu}\,;$$ this provides  the perturbed line element as follows:
\begin{equation}
\label{ds2_T}
\begin{array}{lcl}
ds^2 &=& (q_{\mu\nu}+\delta q_{\mu\nu})dx^{\mu}dx^{\nu} \\[2ex]
&=&\left(1+\fracc{4\alpha\,}{\alpha-3} \delta\Phi\right)\fracc{(dx^0)}{\Z}^2
- \fracc{2\beta}{(\Z)\alpha}\fracc{\delta\Phi,_i}{\Phi_{,0}}\,dx^0dx^i \\[2ex]
&&- \fracc{1}{\alpha}\left(1 +2\,\delta\Phi\right)\,\delta_{ij}\,dx^i dx^j.
\end{array}
\end{equation}
Use of the cosmic time $t$ gives
\begin{eqnarray}
ds^2&=& \left(1-\fracc{4\,\delta\Phi}{3a^2 -1}\right)dt^2 - a^2(1+ 2\delta\Phi)\delta_{ij}\,dx^i dx^j\cr\,\nonumber\\
&&- \fracc{(a^2 -1)(9a^2-1)}{2a^2H}\delta\Phi_{,\,i}\,dx^i dt \label{ds2_t}
\end{eqnarray}
where $H= \dot \Phi$ is the Hubble parameter.

\subsection{Perturbation of the dynamics}
A little calculation shows that
the perturbation of the lhs of the field equation (\ref{dynamic}) is
\begin{equation}
\begin{array}{l}
\delta(\sqrt{V}\Box\Phi)=\,\fracc{a}{2}(3a^2-1)\,\ddot{\delta\Phi }+ \dot a(9a^2-2)\,\dot{\delta\Phi}\, +\\[2ex]
\phantom{\delta(\sqrt{V}\Box\Phi)=}\,- \fracc{(3a^2-1)^3}{8a^3}\Delta \delta\Phi + \\[2ex] \phantom{\delta(\sqrt{V}\Box\Phi)=}\, +\left[3\kappa\chi_0\left(\fracc{3a^2+1}{3a^2-1}\right)-\fracc{6a\dot a^2}{3a^2-1}\right]\delta\Phi\,.
\end{array}
\end{equation}
where $\Delta = \delta^{ij}\partial_i\partial_j$ is the ordinary Laplace operator and $\chi_0$\, is the rhs  of the background  field equation [see Eq.\ \eqref{jr}].

Let us consider the rhs. The unperturbed energy-momentum tensor is that of a perfect fluid
$$T^{\mu}{}_{\nu}=(\rho+p)v^\mu v_\nu-p\,\delta^\mu_\nu,$$
in the comoving coordinates (i.e. with four-velocity  $v^{\mu}=(1,0,0,0)$).
As for the perturbed fluid  the constraint
$$\delta(v^{\mu}v_{\mu})=\,0 ~~\Rightarrow ~~ \delta v^\0=\,-\frac{1}{2}\,\delta q_{\0\0}$$
gives
\begin{eqnarray}
\delta T^{\0}{}{_\0}&=& \delta\rho,\\[1ex]
\delta T^{\0}{}_{i}&=& (\rho+p)\,\delta v_i,\\[1ex]
\delta T^{i}{}_{\0}&=& (\rho+p)\,\delta v^i,\\[1ex]
\delta T^{i}{}_{j}&=& -\,\delta p\,\delta^i_j\,.
\end{eqnarray}
It follows that $\delta T=\delta\rho-3\delta p$ and $\delta E=\delta\rho$. The perturbation of the vector $C_{\mu}$ is given by
\begin{equation}
	\delta C_{\mu} =\, \frac{\beta}{\alpha H^2} \,\left[H \delta T_{\mu}^\0 -\, \delta\rho\,\Phi,_\mu + \,(\rho+p)(v_\mu v^\nu -\, \delta_\mu^\nu)\,\delta\Phi,_\nu\right]\,.\nonumber
\end{equation}
In components,  $\delta C_{t}=0$ and
$$\delta C_{i}=\frac{\beta}{\alpha H^2}(\rho+p)\left(\dot\Phi\,\delta v_i-\delta\Phi,_{\,i}\right).$$
Since $C_\mu$ is zero on the background $\delta C^{\mu}=q^{\mu\nu}\delta C_{\nu}$ and then
$\delta C^{t}=0$ and $\delta C^{i}=-(1/a^2)\,\delta^{ij}\,\delta C_{j}.$ Finally, the perturbation of $C^{\mu}{}_{;\mu}$ is
\begin{equation}
	\delta(C^{\,\mu}{}_{;\mu})=(\delta C^{\,\mu})_{;\mu}= \frac{\beta}{H^2}\, (\rho+p) \left[H\,\delta^{ij}\,\delta v_i,_j - \Delta \delta\Phi\right]
\end{equation}
The energy-momentum conservation equation gives rise to the perturbed continuity equation
\begin{equation}
	\dot{\delta\rho}+\dot{\rho}\,\delta v^t+(1+\lambda)(\Theta\delta\rho+\rho\delta\Theta)=0
\end{equation}
and the perturbed Euler equation\footnote{In the perturbation applied to RG, this equation does not appear explicitly because it is a consequence of the constraint equation $G^t_i=-T^t_i$.}
\begin{equation}
(\rho+p)(\dot{\delta v_{i}}-\delta\Gamma^0_{i0}-\Gamma^0_{ij}\delta v^j-\Gamma^j_{i0}\delta v_j)-\delta p_{,i}+\dot p\,\delta v_i=0,
\end{equation}
where as usual $\Theta = {v^\mu}_{;\mu}$ is the unperturbed expansion factor. Some auxiliary expressions useful
to simplify this equation are
\begin{eqnarray*}
\delta\Gamma^t_{tt}&=&(1/2)\dot{\delta q_{tt}}\,,\\[1ex]
\delta\Gamma^t_{it}&=&(1/2)(\delta q^{tj}q_{ij,t}+\delta q_{tt,i})\,,\\[1ex]
\delta v^i&=&a^{-2}(\delta q_{ti}-\delta v_i).
\end{eqnarray*}

In the end we have four variables ($\delta\rho$ and $\delta v_i$) and four equations which together with the equation for $\delta\Phi$ yield a closed dynamical system. It should be remarked that if more general forms of matter are taken into account (with heat flux or anisotropic pressure components), then the system is not closed anymore assuming only the evolution of the scalar field, the conservation of the energy momentum tensor and the evolution of their perturbations. Similarly to GR, some constitutive equations for the dissipative terms should be added to those ones. Nevertheless, this dynamical system will hardly be overdetermined, since there is no constraint equation for the scalar field in GSG cosmology.

We trade as usual $\delta\rho$ for the contrast density $\delta= \delta\rho/\rho$ and also using the continuity equation of the background we get
\begin{equation}\dot\delta-(1+\lambda)(\Theta\delta v^t-\delta\Theta)=0.\end{equation}
The perturbation of the expansion coefficient is given by
\begin{equation}\delta\Theta=3\dot{\delta\Phi} +\Theta\delta v^t+\delta v^i,_{i} \end{equation}
where we have used
$$\delta(\sqrt{-q})=\sqrt{-q}\left(\frac{9a^2-5}{3a^2-1}\right)\delta\Phi.$$

\subsection{Eigenfunction expansion}
We now expand the perturbation in a complete set of eigenfunctions of the Laplace operator $$\Delta Q=-k^2\, Q;$$
in our flat 3-geometry they  are simply plane waves. This amounts to make a spatial Fourier analysis of the perturbations. The coefficients only depend on time. The coefficients of the $k $-modes are as follows
$$\delta \Phi \to \delta \Phi_k(t)Q_k(x) ,\ \delta\to \delta_k(t)Q_k(x), \  \delta v_i \to V_k(t){Q_k}_{,i}(x).$$ For simplicity the index $k$ will be omitted.
There follows a system of three coupled linear ordinary differential equations:
\begin{widetext}
\begin{eqnarray}
\label{fin_pert_delphi}
\ddot{\delta{\Phi}}&+& 2H\,\frac{(9a^2-2)}{(3a^2-1)}\,\dot{{\delta\Phi}} + \left[6\kappa\frac{\chi_0}{a}\frac{(3a^2+1)}{(3a^2-1)^2}-\frac{12a^2H}{(3a^2-1)^2} + \frac{(3a^2 -1)^2k^2}{4a^4}\right]\delta\Phi \cr	 &=&\,\frac{\kappa\rho_0a^{-(4+3\lambda)}}{(3a^2-1)}\left[\frac{2\,\delta}{3a^2-1} +3\lambda\,\delta -\, \frac{12a^2\delta\Phi}{(3a^2-1)^2}-(1+\lambda)\frac{(a^2-1)(9a^2-1)}{4a^4H^2}(HV-\delta\Phi)k^2\right], \\
\label{fin_pert_velo}
\dot V&+&\frac{2\delta\Phi}{3a^2-1}-\frac{\lambda\delta}{1+\lambda}-3\lambda HV=0,\\
\label{fin_pert_contr}
\dot\delta &+&\, (1+\lambda)\left\{3\,\dot{\delta\Phi} +\, \left[\frac{(a^2-1)(9a^2 -1)}{4a^4H}\, \delta\Phi +\frac{V}{a^2} \right] k^2 \right\}=\,0\,.
\end{eqnarray}
\end{widetext}
Now the question is whether the behaviour of the perturbations is compatible with the possibility of structure formation in GSG cosmology.  To proceed, the explicit time dependence of the background scale factor is needed. In the present  situation $a(t)$ is known analytically only in the decelerated phase. We limit our investigation to that case. {The analysis is simplified by the use of  the conformal time $d\eta=dt/a(t)$. Then, at late times  ($\rho_0$ small and $a\approx \bb \sqrt{\eta} \gg 1$}), Eq.\ (\ref{fin_pert_delphi}) is considerably simplified resulting in
\begin{equation}
\delta\Phi'' +\, \frac{5}{2\eta}\,\delta\Phi' +\frac{9\bb^2}{4}\,k^2\eta\,\delta\Phi=\,0\,.
\end{equation}
In this crude approximation the evolution of $\delta\Phi$ depends neither on the matter content  nor on its perturbations. The general solution is a damped oscillation
\begin{equation}
\label{gen_sol_pert}
\delta\Phi=\frac{\delta\Phi_0\sin(k \bb \eta^{\frac{3}{2}}),}{\eta^{\frac{3}{2}}}
\end{equation}
where $\delta\Phi_0$ is an integration constant.

The simplified equations for the velocity perturbation and the density contrast depend explicitly on the equation of state:
\begin{eqnarray}
\label{l0fin_pert_velo}
&&V'+\frac{2\delta\Phi}{3\bb\sqrt{\eta}}-\frac{\lambda \bb\sqrt{\eta}}{1+\lambda}\, \delta-\frac{3\lambda}{2\eta}V=0 \\
&&\delta' +\, (1+\lambda)\left\{3\,\delta\Phi' +\, \left[\frac{9\bb^2\eta^2}{2}\, \delta\Phi +\frac{V}{\bb \sqrt{\eta}} \right] k^2 \right\}=0.\cr &&
\label{l0fin_pert_contr}
\end{eqnarray}
They can be solved analytically in terms of special functions.  In the case of dust, we get
\begin{equation}
\label{l0fin_pert_velo_sol}
V= V_0-\frac{2}{3\bb}\delta\Phi_0\int{\eta^{-2}\sin(k \bb \eta^{\frac{3}{2}})}d\eta
\end{equation}
and
\begin{widetext}
\begin{eqnarray}
\delta=\delta_0 - \frac{2k^2V_0\,\sqrt{\eta}}{\bb} + 3k\delta\Phi_0\cos(k \bb \eta^{\frac{3}{2}}) + \frac{9}{2}\delta\Phi_0\int{\left[\frac{4k^2}{27 a_\eta^2\sqrt{\eta}} \int^\eta {\frac{\sin(k \bb \bar {\eta}^{\frac{3}{2}})}{\bar{\eta}^2}}d\bar\eta+\frac{\sin(k\bb \eta^{\frac{3}{2}})}{ \eta^{\frac{5}{2}}} -\frac{k \bb}{\eta}\cos(k \bb \eta^{\frac{3}{2}}) \right]}d\eta,\nonumber\\
\label{l0fin_pert_contr_sol}
\end{eqnarray}
\end{widetext}
where $V_0$ and $\delta_0$ are integration constants.


At large scales ($k\eta\ll 1$), the perturbation of the scalar field remains constant $\delta\Phi=\delta\Phi_i(k)$ and the velocity perturbation grows like $V\approx a(\eta) \delta\Phi_i$. The growth of the density contrast  depends on the constant part of the velocity perturbation
\begin{equation}
\delta\approx\delta_0-V_0a(\eta).
\end{equation}

At small scales ($k\eta\gg1$), $\delta\Phi$ has a rapidly decreasing oscillatory behaviour and the velocity perturbation remains almost constant with a small oscillation given by the second term on the rhs of (\ref{l0fin_pert_velo_sol}). On the other hand, the density perturbation grows as follows

\begin{figure}
\begin{center}
\includegraphics[scale=0.45]{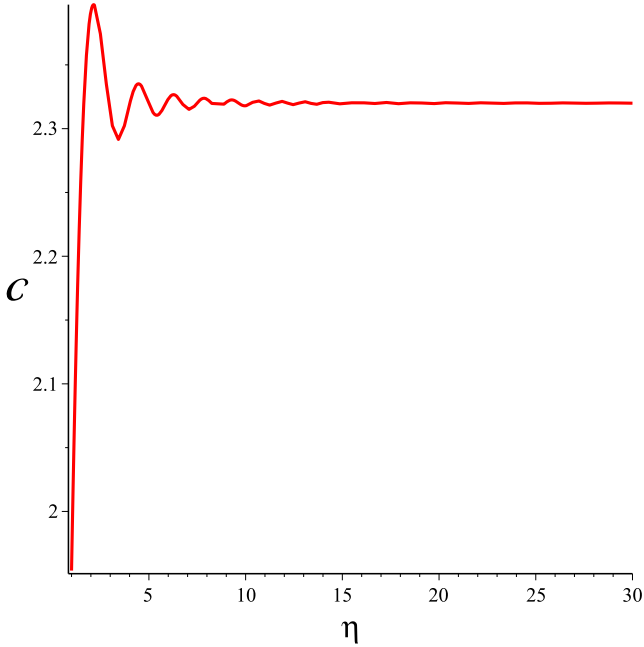}
\end{center}
\caption{Illustrative plot of ${\cal C}(\eta)$ when $k=1$. For large $\eta$, note that ${\cal C}$ tends to a nonzero constant. So, in general, this yields the growing mode of $\delta$ when integrated in $\eta$.}
\label{fig_numc}
\end{figure}

\begin{equation}
\delta\approx \left(\frac{k^2{\cal C}}{3\bb^3}\delta\Phi_0\, - \frac{2k^2V_0}{\bb^2}\right) a(\eta),
\end{equation}
where ${\cal C}(\eta)\equiv\int_\eta{\bar\eta^{-2}\sin(ka_\eta\bar\eta^{\frac{3}{2}})}d\bar\eta$, which has a constant amplitude with a very small oscillation [cf. Fig.\ (\ref{fig_numc})]. The other terms of Eq.\ (\ref{l0fin_pert_contr_sol}) do not contribute in this regime.

The almost empty late time phase of the universe is preceded by a phase that we may try to model phenomenologically as a power law $a(\eta)=a_\eta \eta^m$, where $m>1/2$ is expected to depend on the equation of state. The equations for the perturbations are

\begin{eqnarray}
&& \delta\Phi'' +\, \frac{5m}{\eta}\,\delta\Phi' + \frac{9a_{\eta}^2}{4} \, k^2\eta^{2m}\,\delta\Phi=0, \label{l0fin_pert_phi_rad}\\
&&V' + \frac{2\delta\Phi}{3a_\eta\eta^{m}} - \frac{a_\eta\eta^{m}}{4}\, \delta - \frac{m}{\eta}V=0, \label{l0fin_pert_velo_rad}\\
&&\delta' + 4\,\delta\Phi' + \left[ \frac{3a_\eta^2\eta^{2m+1}}{m} \delta\Phi +\frac{4\,V}{3a_\eta \eta^m} \right] k^2= 0, \quad \label{l0fin_pert_contr_rad}
\end{eqnarray}
where again we are neglecting the contribution from the rhs of Eq.\ (\ref{fin_pert_delphi}).
Setting the integration constant of the decaying mode of $\delta\Phi$ equal to zero, the finite contribution is
\begin{equation}
\label{gen_sol_pert_rad}
\delta\Phi=\delta\Phi_0 \,\eta^{\frac{1-5m}{2}} \, J_{\nu}\left[ \frac{3k\,a_\eta\,\eta^{m+1}}{2(m+1)} \right],
\end{equation}
where $\delta\Phi_0$ is an integration constant and $J_{\nu}(x)$ is a Bessel function of first kind with $\nu=(5m-1)/(2m+2)$. As in the previous case, at large scales $\delta\Phi=\delta\Phi_i(k)$ while at small scales $\delta\Phi$ has a decreasing oscillatory behaviour with power $\eta^{-3m}$.

\begin{figure}
\begin{center}
\includegraphics[scale=0.45]{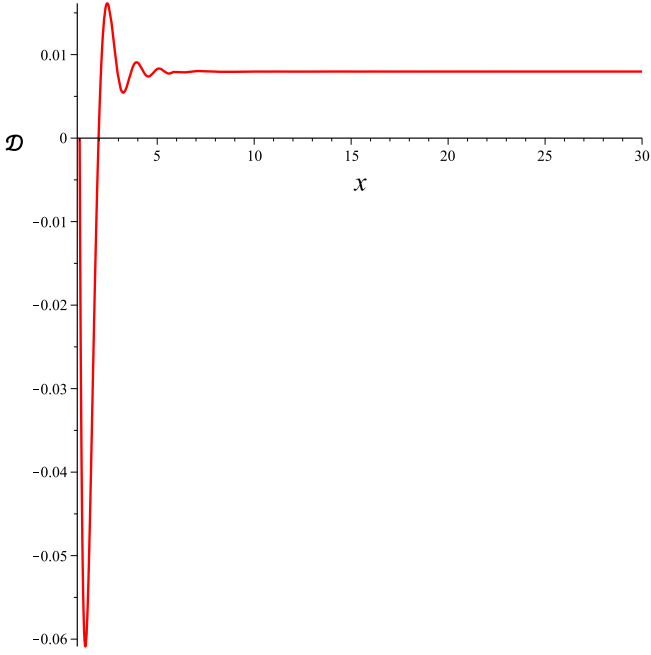}
\end{center}
\caption{Illustrative plot of ${\cal D}(x)$ when $m=3/4$. In comparison to ${\cal C} (\eta)$, the behaviour is analog. This also provides the growing mode for $\delta$ when integrated in time.}
\label{fig_num-d}
\end{figure}

Eliminating $V$ from the equation (\ref{l0fin_pert_contr_rad}), we get the following nonlinear expression for the density contrast
\begin{widetext}
\begin{equation}
\label{eq_delta_rad}
\delta= \delta_0 - 4\,\delta\Phi - \frac{3a_\eta^2\,k^2}{m}\int{\eta^{2m+1}\delta\Phi}d\eta + k^2\int{\int{\left[ \frac{8}{9a_\eta^2}\frac{\delta\Phi}{\bar{\eta}^{2m}} -\frac{\delta(\bar\eta)}{3}\right]}d\bar{\eta}}d\eta = 0.
\end{equation}
\end{widetext}
For $k\eta\ll1$, we may neglect all the terms proportional to $k^2$ and find that $\delta\approx-4\delta\Phi_i$. Using the formula to compute $\delta$ recursively, we can neglect $\delta(\bar\eta)$ in first approximation and see the contribution of the other terms, which is

\begin{equation}
\label{sol_delta_sma_sca}
\delta\propto \delta\Phi_0{\cal D}\,k^2\eta,
\end{equation}
where ${\cal D}(x)\equiv \int_{x}{\bar x^{-5m}\cos(\bar x^{m+1})}d\bar x$ [see its behavior in Fig.\ (\ref{fig_num-d})] and $x$ is equal to the Bessel function's argument in Eq.\ (\ref{gen_sol_pert_rad}). It should be remarked that at small scales, ${\cal D}$ is practically constant and its integration yields the growing mode for $\delta$ whose the rate comparable to $a(\eta)$ will depend specifically on the value of $m$.

\section{Conclusion}
We have discussed cosmology within the context of GSG. The cosmological models have interesting features distinct from those of the corresponding FRW cosmologies. In particular, there is a nonzero special value for the scale factor which cannot be attained by any solution. This fact is responsible for the bounce in the big-universes and the recollapse in the small universes. Some of the main problematic features of the standard cosmological scenario such as the singularity, horizon and flatness problems, can be solved in GSG without making appeal to exotic kinds of matter.

We have also discussed scalar perturbations focusing on the growth of density perturbations in a matter dominated decelerated phase that takes place after an early accelerated phase of the universe and seen that gravitational instabilities are possible in GSG. Fortunately, they are slight different from GR, which means that the GSG cosmology can be tested separately if we interpret properly the data in this new framework. This is left for future work.


\appendix
\section{Gauge transformations in GSG cosmology}\label{A}

In this appendix we show that the first order perturbations of the scalar field
described in section \ref{cosper} give rise to true perturbations of the gravitational metric which cannot arise from infinitesimal coordinate transformations (aka {\em gauge} transformations)
\begin{equation}
x^\alpha \rightarrow \tilde{x}^\alpha=\, x^\alpha +\, \xi^\alpha
\end{equation}
where $\xi^\alpha$ is a ``small" space-time displacement.
To verify the above assertion one has to compare the perturbed metric (\ref{ds2_t}) with the gauge transformed metric
\begin{equation}
\label{met}
q_{\mu\nu} + \delta q_{\mu\nu}=q_{\mu\nu}-q_{\mu\nu},_\gamma\,\xi^\gamma -q_{\mu\gamma}\,\xi^\gamma,_\nu - \,q_{\gamma\nu}\,\xi^\gamma,_\mu\
\end{equation}
taking into account that the scalar field transforms under coordinate transformations as follows
\begin{equation}
\delta\Phi=\, \tilde{\Phi}\,(x) - \Phi\,(x) =\, -\,\xi^\alpha\,\Phi,_\alpha = -H \xi^\0. \label{aa}
\end{equation}
Eq.\ (\ref{met}) and (\ref{aa}) imply the following relations
\begin{equation}
\frac{(\delta\Phi/H)\,\dot{}}{\delta\Phi/H} =\, -\,\frac{2}{3a^2 -1}\,H\,,\label{e1}
\end{equation}
\begin{equation}
\dot{\xi}^i=\,-\,\frac{(3a^2-1)}{4a^4H}^2\,\delta\Phi,_i\,,\label{e2}
\end{equation}
\vskip5pt
\begin{equation}
\xi^i,_j + \, \xi^j,_i = 0 . \label{e3}\end{equation}
The first  equation gives the  time dependence of the perturbations generated by $\xi^{\mu}$
\begin{equation}
\label{pert}
\delta\Phi(t, \vec{x})=\, \frac{a^2H}{3a^2-1}\,e^{\,g(\vec{x})}
\end{equation}
where $g(\vec{x})$ is an arbitrary function of the spatial coordinates. Substituting this in Eqs.\ (\ref{e2}) and \eqref{e3} we can obtain the spatial components of $\xi^{\mu}$.
However, a straightforward calculation shows that  the time dependence of $\delta\Phi$ given in \eqref{pert} is incompatible with Eqs.\ (\ref{fin_pert_delphi}-\ref{fin_pert_contr}). Therefore a solution of such equations cannot be a gauge artifact.

\end{sloppypar}

\section*{Acknowledgements}
We would like to thank the participants of the Spontaneous Workshop VIII (Carg\`ese, 2014) for fruitful discussions. This work was partially supported by CNPq, FAPERJ and the CAPES-ICRANet program (BEX 13956/13-2).


\begin{thebibliography}{100}
\bibitem{nord}
G. Nordstr\"om, {\em Phys.\ Zeit.} {\bf 13} 1126 (1912).
\bibitem{entwurf}
A. Einstein and M. Grossmann, {\em Zeitschrift f\"ur Mathematik und Physik} {\bf 62} 225-261 (1913).
\bibitem{deruelle}
N. Deruelle, {\em Gen.\ Rel.\ Grav.} {\bf 43} 3337 (2011); [gr-qc/1104.4608]
\bibitem{giulini}
D. Giulini, {\em Studies in History and Philosophy of Modern Physics} {\bf 39} 154 (2008); [gr-qc/0611100v2].
\bibitem{JCAP}
M. Novello, E. Bittencourt, U. Moschella, E. Goulart, J.M. Salim and J.D. Toniato, {\em J. Cosmol. Astropart. Phys.}, JCAP06(2013)014.
\bibitem{feynman}
R.P. Feynman, F.B. Morinigo and W.G. Wagner, \textit{Feynman\ lectures on gravitation}, Addison Wesley Pub. Company, Massachusetts, (1995).
\bibitem{novelloetal}
M. Novello and E. Goulart, {\em Class.\ Quantum\ Grav.} {\bf 28} 145022 (2011).
\bibitem{novelloetal2}
E. Goulart, M. Novello, F.T. Falciano and J.D. Toniato, {\em Class.\ Quantum\ Grav.} {\bf 28} 245008 (2011).
\bibitem{einstein}
A. Einstein, \textit{The meaning of relativity}, Princeton University Press, (1950).
\bibitem{narlikar}
J.V. Narlikar, \textit{Introduction to Cosmology}, (Cambridge University Press, Great Britain, 1993)
\bibitem{mukhanov}
V. Mukhanov, \textit{Physical Foundations of Comsology}, (Cambridge University Press, New York, 1973)
\bibitem{haw_ell}
S.W. Hawking and G.F.R. Ellis, \textit{The Large Scale Structure of the Space-Time} (Cambridge University Press, New York, 2005).
\bibitem{novellobergli}
M. Novello and S.E.P. Bergliaffa, {\em Physics Reports} {\bf 463} 4 (2008).
\bibitem{lifshitz}
E. Lifshitz, {\em Sov.\ Phys.\ JETP} {\bf 10} 116 (1946).
\bibitem{padma}
T. Padmanabhan, \textit{Gravitation: Foundations and Frontiers}, Cambridge University Press, Cambridge (2010).
\end{thebibliography}
\end{document}